\documentstyle[12pt]{article}
\author{Igor G. Korepanov}
\title{The tetrahedral analog of Veneziano amplitude}
\date{}

\def\be{\begin{equation}}
\def\ee{\end{equation}}

\def\hitrayakomanda#1 #2 #3 #4 #5 #6{\pmatrix{a\vphantom a_{#1}^{#6} &
b\vphantom b_{#1}^{#6}\cr \skipchik
c\vphantom c_{#1}^{#6} & d\vphantom d_{#1}^{#6} }
& \mapsto & \pmatrix{\tilde a\vphantom a_{#1}^{#6} &
\tilde b\vphantom b_{#1}^{#6}\cr \skipchik
\tilde c\vphantom c_{#1}^{#6} & \tilde d\vphantom d_{#1}^{#6}} = 
\pmatrix{ #2_d^{-1} & 0 \cr \skipchik  0 & #3_d^{-1}}
\pmatrix{a\vphantom a_{#1}^{#6} & b\vphantom b_{#1}^{#6}\cr
\skipchik
c\vphantom c_{#1}^{#6} & d\vphantom d_{#1}^{#6} }
\pmatrix{ #4_d & 0 \cr \skipchik 0 & #5_d }  }

\def\skipchik{\noalign{\vskip0.05\baselineskip}}

\def\pder#1#2{{\partial#1\over\partial#2}}

\def\d{{\rm d}}

\begin{document}
\maketitle

\begin{abstract}
In {\tt solv-int/9812016} it was shown that the Veneziano amplitude
in string theory comes naturally from one of the simplest solutions
of the functional pentagon equation (FPE). More generally, FPE is
intimately connected with the duality condition for scattering
processes. Here I find the amplitude that comes the same way from
a solution of the functional tetrahedron equation, with
the duality replaced by the local Yang--Baxter equation.
\end{abstract}

\section{Introduction}

It was shown in~\cite{kor sai} that the famous Veneziano amplitude, from
which all the string theory starts, comes naturally from one of the
simplest solutions of the functional pentagon equation (FPE).
More generally, FPE is
intimately connected with the duality condition for scattering
processes.

 From the viewpoint of the theory of integrable models, FPE is a rather
trivial equation whose solutions have transparent geometrical or
group-theoretic meaning~\cite[section~5]{kor sai}. It looks natural
to search for similar constructions with FPE replaced by
the functional tetrahedron equation (FTE). As the relations between
the pentagon and duality condition are like those between the tetrahedron
and local Yang--Baxter equation (LYBE), the duality condition is likely
to be replaced by LYBE.

In this paper, I find such FTE and LYBE solutions that are described
by formulas very similar to those describing Veneziano amplitude
in~\cite{kor sai}, including the fundamental property of M\"obius
invariance. They are what I mean by the tetrahedral analog of
Veneziano amplitude.

\section{A functional transformation for edge variables from
refactorization equation}

Consider the following ``refactorization equation''
for the product of three matrices:
\begin{eqnarray}
&&
\pmatrix{ a_1&b_1& 0\cr c_1&d_1& 0 \cr 
0 & 0 & 1 }
\pmatrix{ a_2& 0 &b_2\cr 0 & 1 & 0 \cr 
c_2& 0 &d_2 }
\pmatrix{ 1 & 0 & 0 \cr 
0 &a_3&b_3\cr 0 &c_3&d_3 } = 
\nonumber\\[0.5\normalbaselineskip]
&& =
\pmatrix{ 1 & 0 & 0 \cr 
0 &a'_3&b'_3\cr 0 &c'_3&d'_3 }
\pmatrix{ a'_2& 0 &b'_2\cr 0 & 1 & 0 \cr 
c'_2& 0 &d'_2 }
\pmatrix{ a'_1&b'_1& 0 \cr c'_1&d'_1& 0 \cr
0 & 0 & 1 },
\label{ieq pererazlozhenie}
\end{eqnarray}
($a_1,\ldots, d'_3$ are numbers) for the case when all six
submatrices $\pmatrix{a_i^{(\prime)} & b_i^{(\prime)}\cr
\skipchik
c_i^{(\prime)} & d_i^{(\prime)}}$ have the form
\be
\pmatrix{a & b\cr c & d}=\pmatrix{\alpha & 1-\alpha\cr 1-\beta & \beta}.
\label{eq alpha}
\ee
In other words, each of the six matrices in~(\ref{ieq pererazlozhenie})
transforms the vector $\pmatrix{1\cr 1\cr 1}$ into itself.

It is known from~\cite{pomi-iv,dokt,troe} that each side of
(\ref{ieq pererazlozhenie}) determines the other side to within
some ``gauge freedom'', and one can verify that the additional
conditions~(\ref{eq alpha}) are exactly good for fixing that freedom.

The fate of an {\em arbitrary\/}
vector $\pmatrix{p\cr q\cr r}$ under the action
of both sides of (\ref{ieq pererazlozhenie}) is more
complicated. We present it in Figure~\ref{fig pqr},
\begin{figure}[hb]
\begin{center}
\unitlength=1mm
\special{em:linewidth 0.4pt}
\linethickness{0.4pt}
\begin{picture}(100.00,40.00)
\put(0.00,0.00){\line(1,1){40.00}}
\put(30.00,40.00){\line(0,-1){40.00}}
\put(0.00,10.00){\line(1,0){40.00}}
\put(3.00,4.00){\makebox(0,0)[rb]{$y$}}
\put(3.00,11.00){\makebox(0,0)[cb]{$x$}}
\put(29.00,0.00){\makebox(0,0)[rb]{$z$}}
\put(38.00,11.00){\makebox(0,0)[cb]{$p$}}
\put(36.00,37.00){\makebox(0,0)[rb]{$q$}}
\put(29.00,40.00){\makebox(0,0)[rt]{$r$}}
\put(23.00,11.00){\makebox(0,0)[cb]{$u$}}
\put(20.00,21.00){\makebox(0,0)[rb]{$v$}}
\put(31.00,17.00){\makebox(0,0)[lc]{$w$}}
\put(11.00,9.00){\makebox(0,0)[lt]{$X_1$}}
\put(31.00,9.00){\makebox(0,0)[lt]{$X_2$}}
\put(31.00,29.00){\makebox(0,0)[lt]{$X_3$}}
\put(30.00,30.00){\circle*{1.00}}
\put(30.00,10.00){\circle*{1.00}}
\put(10.00,10.00){\circle*{1.00}}
\put(60.00,30.00){\line(1,0){40.00}}
\put(70.00,40.00){\line(0,-1){40.00}}
\put(60.00,0.00){\line(1,1){40.00}}
\put(70.00,30.00){\circle*{1.00}}
\put(70.00,10.00){\circle*{1.00}}
\put(90.00,30.00){\circle*{1.00}}
\put(60.00,31.00){\makebox(0,0)[lb]{$x$}}
\put(62.00,3.00){\makebox(0,0)[rb]{$y$}}
\put(69.00,0.00){\makebox(0,0)[rb]{$z$}}
\put(100.00,31.00){\makebox(0,0)[rb]{$p$}}
\put(96.00,37.00){\makebox(0,0)[rb]{$q$}}
\put(71.00,40.00){\makebox(0,0)[lt]{$r$}}
\put(80.00,31.00){\makebox(0,0)[cb]{$f$}}
\put(79.00,20.00){\makebox(0,0)[rb]{$g$}}
\put(69.00,20.00){\makebox(0,0)[rc]{$h$}}
\put(69.00,31.00){\makebox(0,0)[rb]{$Y_2$}}
\put(91.00,29.00){\makebox(0,0)[lt]{$Y_1$}}
\put(69.00,11.00){\makebox(0,0)[rb]{$Y_3$}}
\end{picture}
\end{center}
\caption{}
\label{fig pqr}
\end{figure}
where we denote the matrices entering~(\ref{ieq pererazlozhenie}), in their
order in that equation, by letters $X_1$, $X_2$, $X_3$, $Y_3$, $Y_2$,
$Y_1$. The meaning of the LHS of Figure~\ref{fig pqr} is that
$$
X_3 \pmatrix{p\cr q\cr r} = \pmatrix{p\cr v\cr w}, \qquad
X_2 \pmatrix{p\cr v\cr w} = \pmatrix{u\cr v\cr z}, \qquad
X_1 \pmatrix{u\cr v\cr z} = \pmatrix{x\cr y\cr z},
$$
while the meaning of the RHS is that
$$
Y_1 \pmatrix{p\cr q\cr r} = \pmatrix{f\cr g\cr r}, \qquad
Y_2 \pmatrix{f\cr g\cr r} = \pmatrix{x\cr g\cr h}, \qquad
Y_3 \pmatrix{x\cr g\cr h} = \pmatrix{x\cr y\cr z}.
$$

One can see that if, vice versa, all the values $x,y,z,\ldots$
in e.g.\ the LHS of Figure~\ref{fig pqr} are given, then matrices
$X_1,X_2,X_3$
of the form~(\ref{eq alpha}) are recovered unambiguously.
So, we can take some given values of nine numbers in the LHS,
get the triple of matrices $X_1,X_2,X_3$ from them, then get
$Y_1,Y_2,Y_3$ by (\ref{ieq pererazlozhenie}), and then get the
missing values $f,g,h$ in the RHS from $p,q,r$ using $Y_1,Y_2,Y_3$.
We will formulate this the following way: for any fixed ``outer''
variables $x$, $y$, $z$, $p$, $q$, $r$, the transformation
\be
R=R(x,y,z,p,q,r)\;\colon\;\; (u,v,w)\mapsto (f,g,h)
\label{eq R}
\ee
is given.

The transformations (\ref{eq R}) satisfy the {\em functional
tetrahedron equation\/} (FTE).
To explain this, note that equation (\ref{ieq pererazlozhenie})
can be naturally regarded as an equation in the direct sum of three
one-dimensional complex linear spaces, each of the matrices
acting nontrivially only in a direct sum of two
of them. One can consider similar relations in a direct sum of {\em four\/}
spaces (each of the matrices acting nontrivially again only in a direct sum
of two spaces). Let us picture in Figure~\ref{fig 4lines}
\begin{figure}[ht]
\begin{center}
\unitlength=1mm
\special{em:linewidth 0.4pt}
\linethickness{0.4pt}
\begin{picture}(100.00,40.00)
\put(0.00,5.00){\line(2,1){40.00}}
\put(0.00,10.00){\line(1,0){40.00}}
\put(30.00,0.00){\line(0,1){40.00}}
\put(15.00,0.00){\line(1,2){20.00}}
\put(60.00,30.00){\line(1,0){40.00}}
\put(70.00,40.00){\line(0,-1){40.00}}
\put(65.00,0.00){\line(1,2){20.00}}
\put(60.00,15.00){\line(2,1){40.00}}
\put(10.00,10.00){\circle*{1.00}}
\put(20.00,10.00){\circle*{1.00}}
\put(30.00,10.00){\circle*{1.00}}
\put(23.33,16.67){\circle*{1.00}}
\put(30.00,20.00){\circle*{1.00}}
\put(30.00,30.00){\circle*{1.00}}
\put(70.00,30.00){\circle*{1.00}}
\put(80.00,30.00){\circle*{1.00}}
\put(90.00,30.00){\circle*{1.00}}
\put(70.00,10.00){\circle*{1.00}}
\put(70.00,20.00){\circle*{1.00}}
\put(76.67,23.33){\circle*{1.00}}
\put(9.00,11.00){\makebox(0,0)[rb]{$1$}}
\put(21.00,9.00){\makebox(0,0)[lt]{$2$}}
\put(31.00,9.00){\makebox(0,0)[lt]{$4$}}
\put(31.00,19.00){\makebox(0,0)[lt]{$5$}}
\put(29.00,31.00){\makebox(0,0)[rb]{$6$}}
\put(22.50,17.90){\makebox(0,0)[rb]{$3$}}
\end{picture}
\end{center}
\caption{}
\label{fig 4lines}
\end{figure}
the spaces as straight lines, put matrices at their intersections,
and attach the results of matrix action upon some
4-vector to line segments like in Figure~\ref{fig pqr},
and then consider the transition from the LHS of Figure~\ref{fig 4lines}
to its RHS as a composition of ``elementary'' transformations~$R$
of type~(\ref{eq R}).

As was explained in the paper~\cite{troe} (and the reader will verify
it him-/herself easily), there exist two different compositions of four
$R$s both transforming the LHS of Figure~\ref{fig pqr} in its RHS.
The first of them starts with~$R_{356}$, by which we mean ``turning
inside out'' triangle~$356$, while the other---with~$R_{123}$.
We can write FTE in the same abstract form as in~\cite{troe}:
\be
R_{123} \circ R_{145} \circ R_{246} \circ R_{356} =
R_{356} \circ R_{246} \circ R_{145} \circ R_{123},
\label{ieq1}
\ee
but the sense of (\ref{ieq1}) is now different: $R$ is now a transformation
of variables belonging to the {\em edges\/} rather than of matrices
belonging to vertices.

To prove FTE (\ref{ieq1}) for edge variables, note that the variables
belonging to {\em inner\/} edges (i.e., say, edges $12$, $13$, \ldots,
$56$ in the LHS of Figure~\ref{fig 4lines}) are unambiguously recovered
if variables at {\em outer\/} edges and matrices at vertices are given.
The FTE for matrices, according to~\cite{troe}, does hold, while the
variables at outer edges are not changed by the transformations. Thus,
the variables at inner edges do not depend on the way of transformations
as well.

\section{M\"obius invariance}

The same way as we have traced the fate of vector $\pmatrix{p\cr q\cr r}$
under the action of LHS and RHS of (\ref{ieq pererazlozhenie})
in Figure~\ref{fig pqr}, we can trace the fate of two more vectors,
namely
\begin{equation}
\pmatrix{p_n\cr q_n\cr r_n} = \kappa \pmatrix{1\cr 1\cr 1} +
\lambda \pmatrix{p\cr q\cr r}
\quad \hbox{and} \quad
\pmatrix{p_d\cr q_d\cr r_d} = \mu \pmatrix{1\cr 1\cr 1} +
\nu \pmatrix{p\cr q\cr r},
\label{*}
\end{equation}
where $\kappa,\lambda,\mu,\nu$ are some constants (and subcripts
$n$ and $d$ stand for ``numerator'' and ``denominator'', see
formula~(\ref{eq ed}) below).
I do not draw here corresponding diagrams, differing from
Figure~\ref{fig pqr} only in that $n$ or $d$ is added to all small
letters.

Now let us do the following {\em gauge transformations\/} (in the
sense of~\cite{pomi-iv,dokt,troe}) on matrices $X_1,\ldots,Y_3$:
\begin{eqnarray*}
\hitrayakomanda 1 x y u v {}, \\
\hitrayakomanda 2 u z p w {}, \\
\hitrayakomanda 3 v w q r {}, \\
\hitrayakomanda 1 f g p q {\prime}, \\
\hitrayakomanda 2 x h f r {\prime}, \\
\hitrayakomanda 3 y z g h {\prime}.
\end{eqnarray*}
Here, of course, $x_d=\mu+\nu x$ etc., in analogy with $p_d$, $q_d$ and~$r_d$ in~(\ref{*}).

Denote the so obtained matrices $\tilde X_1, \ldots, \tilde Y_3$. Now
imagine a version of Figure~\ref{fig pqr} for these matrices with tildes.
One can say that the transformation of vectors corresponding to the
above gauge matrix transformation has brought all variables with
subscript~$d$ into~$1$, and hence the matrices with tildes have again
the form~(\ref{eq alpha}). As for the variables with subscript~$n$,
they turned into
\be
x\to \tilde x={x_n\over x_d}={\kappa+\lambda x\over \mu+\nu x}
\quad \hbox{etc.}
\label{eq ed}
\ee

We see from here that a linear-fractional (M\"obius) transformation
of variables $x,y,\ldots$ commutes with the transformation~$R$.
Clearly, the same conclusion could be made from the explicit formulas
for $R$ given in Section~\ref{sec explicit}. The M\"obius invariance
is an argument in support of the idea that $R$ really is an analog of
``pentagonal'' transformation from~\cite{kor sai} connected with
Veneziano amplitude.

\section{Connection between volume elements and the explicit form of
functional transformation}
\label{sec explicit}

Let us now vary the edge variables in Figure~\ref{fig pqr}, with
matrices $X_1,\ldots Y_3$ fixed. For instance, consider the variables
at outer edges of the LHS of that Figure as functions of three inner
variables $u,v,w$, and calculate the corresponding partial derivatives.
The reader will easily check that
\be
\pder xu={x-v\over u-v},\quad
\pder xv={x-u\over v-u},\quad
\pder yu={y-v\over u-v}
\label{eq pder}
\ee
and so on.

Using formulas of the type (\ref{eq pder}), it is not hard to obtain
the following relations for ``volume elements'':
\be
\d x \wedge \d y \wedge \d z =
{x-y\over u-v}{z-u\over w-u} \,\d u \wedge \d v \wedge \d w
\label{eq diff1}
\ee
 from the LHS of Figure~\ref{fig pqr} and similarly
\be
\d x \wedge \d y \wedge \d z =
{x-h\over f-h}{y-z\over g-h} \,\d f \wedge \d g \wedge \d h
\label{eq diff2}
\ee
 from its RHS. The equalness of the RHSs of (\ref{eq diff1}) and
(\ref{eq diff2}) can be called ``the relation between
$\d u \wedge \d v \wedge \d w$ and $\d f \wedge \d g \wedge \d h$
got via $\d x \wedge \d y \wedge \d z$''.

Similarly, the equalness of the RHSs of relations
\be
\d y \wedge \d z \wedge \d p =
{y-u\over v-u}{z-p\over w-u} \,\d u \wedge \d v \wedge \d w
\label{eq diff3}
\ee
and
\be
\d y \wedge \d z \wedge \d p =
{y-z\over g-h}{p-g\over f-g} \,\d u \wedge \d v \wedge \d w
\label{eq diff4}
\ee
can be called ``the relation between
$\d u \wedge \d v \wedge \d w$ and $\d f \wedge \d g \wedge \d h$
got via $\d y \wedge \d z \wedge \d p$''. There are four more pairs
of relations of the type (\ref{eq diff1}--\ref{eq diff4}) with
$\d z \wedge \d p \wedge \d q$, $\d p \wedge \d q \wedge \d r$,
$\d q \wedge \d r \wedge \d x$ and $\d r \wedge \d x \wedge \d y$
respectively in their LHSs.

Certainly, one can exclude the differentials from those relations
and obtain formulas giving explicitely the connection between
edge variables, i.e.\ the transformation~$R$, namely
\begin{eqnarray}
{x-y\over u-y}\,{u-z\over p-z}&=&{x-h\over f-h}\,{f-g\over p-g},
\label{eqnarray nachalo} \\
{y-x\over v-x}\,{v-r\over q-r}&=&{y-h\over g-h}\,{g-f\over q-f}, \\
{z-p\over w-p}\,{w-q\over r-q}&=&{z-g\over h-g}\,{h-f\over r-f}, \\
{x-v\over u-v}\,{u-w\over p-w}&=&{x-r\over f-r}\,{f-q\over p-q}, \\
{y-u\over v-u}\,{v-w\over q-w}&=&{y-z\over g-z}\,{g-p\over q-p}, \\
{z-u\over w-u}\,{w-v\over r-v}&=&{z-y\over h-y}\,{h-x\over r-x}.
\label{eqnarray konec}
\end{eqnarray}

\section{Local Yang--Baxter equation}

The local Yang--Baxter equation (LYBE) dealt with in this section
differs from the conventional Yang--Baxter equation, first, in its
continuous (instead of usual discrete) ``set of colours'' and, second
(and this is what makes it ``local'', or ``twisted''),
in that all six $R$-matrices (instead of which we will have, however,
functions of 4 complex variables) entering it are different
(in the usual Yang--Baxter equation, the LHS and RHS are made from the same
3 matrices, multiplied in different orders). Namely, our LYBE will
have the following form (for {\em real\/} $x,y,\ldots$):
\begin{eqnarray}
\int L(x,y,u,v)\, M(u,z,p,w)\, N(v,w,q,r) \,\d u \wedge \d v \wedge \d w
\nonumber\\
=\int N'(y,z,g,h)\,M'(x,h,f,r)\,L'(f,g,p,q) \,\d f \wedge \d g \wedge \d h.
\label{eq int}
\end{eqnarray}

In the same way as the duality relation in~\cite{kor sai}, the
equality~(\ref{eq int}) will hold if we require that the relation
hold obtained from (\ref{eq int}) by {\em removing the integration
signs}, with the triples of variables $u,v,w$ and $f,g,h$
connected by some dependence. For such dependence, we will take
the transformation $R$ from formula~(\ref{eq R}). Then, the following
construction of functions $L,\ldots,N'$ can be proposed.

Take the relation
\be
{x-y\over u-v}{z-u\over w-u} \,\d u \wedge \d v \wedge \d w =
{x-h\over f-h}{y-z\over g-h} \,\d f \wedge \d g \wedge \d h
\label{eq diff}
\ee
(see (\ref{eq diff1}, \ref{eq diff2})), and also the relations
(\ref{eqnarray nachalo}--\ref{eqnarray konec}) raised in arbitrary
degrees (the relations (\ref{eqnarray nachalo}--\ref{eqnarray konec})
are not independent, so one of those degrees can be set to zero).
Then multiply separately the LHSs and RHSs of all so obtained relations
(including (\ref{eq diff})). The obtained LHS and RHS will be exactly
the integrands in (\ref{eq int}), and from them the multipliers
$L,\ldots,N'$ depending on proper quadruples of variables are easily
extracted.

I leave for further work the problem of possible choices of integration
domains in~(\ref{eq int}) and integral regularization (if needed).
The explicit form of functions $L,\ldots,N'$ will also be presented
elsewhere. Let me just note that we can also regard all
the variables $x,y,\ldots$ as {\em complex}. In such case, we should
multiply the integrands (including differentials) by their complex
conjugates and integrate over some domains of {\em six\/} real dimensions.
This will be the tetrahedral analog of Virasoro--Shapiro amplitude.

\section{Discussion}

The LYBE of the form (\ref{eq int}), as well as the duality
equations from~\cite{kor sai}, are interesting because there
exists a hope to construct from them interesting ``exactly solvable''
functional integrals, perhaps connected with 3-dimensional statistical
physics. By the way, here I have presented the tetrahedral analog of
{\em one\/} of two models in~\cite{kor sai}, and it seems fascinatingly
interesting to construct the analog of the other model (and their
generalizations). Very interesting will be also to clarify the relations
between pentagon and tetrahedron equations, where, despite the presence
of the excellent work~\cite{kash ser}, many things are unclear.

\subsection*{Acknowledgements}

I am grateful to Satoru Saito for valuable discussions during my stay
at Tokyo Metropolitan University, in the course of which the idea started
to revisit, from the modern integrability theory viewpoint, the algebraic
structures from which string theory was born in its time. I am also
grateful to Sergei Sergeev for many discussions on tetrahedron and
pentagon equations. Finally, I am glad to thank Russian Foundation
for Basic Research for its (mostly moral) support under
grant no.~98-01-00895.


\begin{thebibliography}{99}

\bibitem{kor sai}
I.G. Korepanov and S. Saito,
{\sl Finite-dimensional analogs of string $s \leftrightarrow t$
duality and pentagon equation},
solv-int/9812016,
accepted for publication in Theor. Math. Phys. (1999).

\bibitem{pomi-iv}
I.G. Korepanov,
{\sl A dynamical system connected with inhomogeneous 6-vertex model},
Zapiski Nauch. Semin. POMI {\bf 215}, 178--196 (1994),
also available as hep-th/9402043.

\bibitem{dokt}
I.G. Korepanov,
{\sl Integrable systems in discrete space--time and inhomogeneous models
of two-dimensional statistical physics},
Thesis for obtaining the ``doktor nauk" scientific degree,
161~pp., S-Petersburg, 1995 (in Russian. English version is
available as solv-int/9506003).

\bibitem{troe}
R.M.~Kashaev, I.G.~Korepanov and S.M.~Sergeev,
{\sl Functional Tetrahedron Equation},
Teor. Mat. Fiz. {\bf 117}:3, 370--384 (1998), also solv-int/9801015.

\bibitem{kash ser}
R.M.~Kashaev and S.M.~Sergeev,
{\sl On pentagon, ten-term, and tetrahedron relations},
Comm. Math. Phys. {\bf 195}, 309--319 (1998), also q-alg/9607032.


\end{thebibliography}
\end{document}